\documentclass[pdflatex,sn-nature]{sn-jnl}

\usepackage{graphicx}%
\usepackage{multirow}%
\usepackage{amsmath,amssymb,amsfonts}%
\usepackage{amsthm}%
\usepackage{mathrsfs}%
\usepackage[title]{appendix}%
\usepackage{xcolor}%
\usepackage{textcomp}%
\usepackage{manyfoot}%
\usepackage{booktabs}%
\usepackage{algorithm}%
\usepackage{algorithmicx}%
\usepackage{algpseudocode}%
\usepackage{listings}%
\usepackage{hyperref}
\usepackage{braket}
\usepackage{comment}

\theoremstyle{thmstyleone}%
\theoremstyle{thmstyletwo}%
\usepackage{gensymb}
\theoremstyle{thmstylethree}%

\begin{document}
	
	\title[Article Title]{Generation of bright quantum high-order harmonic driven by combined coherent and bright squeezed vacuum light}
	
	\author[1,2]{\fnm{Wentao} \sur{Wang}}
	\equalcont{These authors contributed equally to this work.}
	
	\author[1,2]{\fnm{Yaoshun} \sur{Sun}}
	\equalcont{These authors contributed equally to this work.}
	
	\author[1,2]{\fnm{Liyuan} \sur{Wang}}
	
	\author[3]{\fnm{Lingrui} \sur{Hu}}
	
	\author[1,2]{\fnm{Dajun} \sur{Ding}}
	
	\author*[1,2]{\fnm{Xiangyu} \sur{Tang}}\email{tangxiangyu@jlu.edu.cn}
	
	\author*[1,2]{\fnm{Mingxuan} \sur{Li}}\email{mingxuanli@jlu.edu.cn}
	
	\author*[1,2]{\fnm{Jianmin} \sur{Yuan}}\email{yuanjianmin@jlu.edu.cn}
	
	\author*[1,2]{\fnm{Sizuo} \sur{Luo}}\email{luosz@jlu.edu.cn}
	
	\affil[1]{\orgdiv{Institute of Atomic and Molecular Physics}, \orgname{Jilin University}, \city{Changchun}, \postcode{130012}, \country{China}}
	
	\affil[2]{\orgdiv{Jilin Provincial Key Laboratory of Transient Quantum Process Control and Application}, \orgname{Jilin University}, \city{Changchun}, \postcode{130012}, \country{China}}
	
	\affil[3]{\orgdiv{College of Physics}, \orgname{Jilin University}, \city{Changchun}, \postcode{130012}, \country{China}}
	\abstract{Attosecond quantum light, formed by the superposition of high-order harmonics driven by intense quantum light, opens new routes to probe quantum-mechanical correlations in matter. In this study, we have investigated the macroscopic propagation effects of quantum high-order harmonics generated by the combination of strong coherent and weak bright squeezed vacuum (BSV) lasers interacting with atomic gas. Our results reveal that the pressure-dependent intensity of harmonics arising from absorbing or emitting BSV photons differs from that of harmonics generated using only strong coherent pulses. Macroscopic propagation simulations indicate that the action phase of harmonics is perturbed by the weak BSV pulses. This perturbation modulates the phase mismatch of sub-cycle attosecond bursts and affects their quantum properties when the gas pressure varies. The ability to generate bright quantum high-order harmonics lays a foundation for the establishment and application of attosecond quantum spectroscopy.}
	
	
	\maketitle
	\section*{Introduction}\label{sec1}
	Attosecond pulses are a unique tool for tracing electron dynamics of matters in real time. They are typically generated by driving gases, liquids, plasmas, or solids with intense coherent lasers through an extreme nonlinear up-conversion process known as high-harmonic generation (HHG)\cite{Anne1993,Luu2018,Ghimire2018,Ganeev2007}. The mechanism can be interpreted as the field-induced semiclassical motion of an electron wave packet in a strong coherent electric field\cite{Corkum1993,Lewenstein1994}. The classical properties of attosecond pulses have been extensively studied and form the foundation of attosecond spectroscopy\cite{Krausz2009,Kienberger2004,Hentschel2001,LiPRL2025Photoionization,li_SA2025CO2}. Building on these developments, recent efforts have explored the quantum properties of attosecond pulses driven by coherent\cite{Bhattacharya2023,Tsatrafyllis2019,Lewenstein2021,Yi2025,Gorlach2023,Lemieux2025} or quantum light sources\cite{Gorlach2023,Lemieux2025,Heimerl2025,Tzur2025,RiveraDean2026,Heimerl2024,Sennary2025,Sennary2026}, aiming to investigate the transfer of quantum states during HHG and to establish attosecond quantum spectroscopy. Theoretically, strong-field processes—including HHG, above-threshold ionization, double ionization, and molecular fragmentation driven by quantum infrared fields—have been predicted and investigated, revealing how quantum fluctuations and correlations of the driving field can influence electron dynamics and be transferred to the emitted radiation\cite{Liu2026a,Stammer2023,RiveraDean2025,Tzur2024,Lyu2025,Liu2025,Long2025,Wang2024,Fang2023,RiveraDean2025a,Stammer2025,Li2026}. These studies highlight the potential of probing quantum behavior under extreme nonlinear conditions and generating XUV photons with quantum properties on the attosecond timescale.
	
	Recently, intense quantum light sources such as bright squeezed vacuum (BSV) have enabled direct driving of tunneling ionization from atoms and nanotips as well as HHG in condenses matter, where the quantum properties of the emitted electrons and photons have been experimentally investigated\cite{Lemieux2025,Heimerl2024,Jiang2026, Liu2026, Rasputnyi2024}. The strong-field ionization by a BSV beam has also been experimentally investigated, demonstrating the boosting and control of tunneling ionization\cite{Jiang2026}, and the enhancement of spider-like
	holographic structures in photoelectron momentum distributions\cite{Liu2026}. Moreover, the investigation on motion of charged particles in BSV field find that BSV induces width oscillations, akin to
	electron quivering in laser light, with an equivalent ponderomotive energy\cite{EvenTzur2024}. An alternative approach combines a strong coherent laser with a weak BSV field, which has the potential to extend such studies to atomic gases with higher ionization potentials while enabling the generation of XUV photons carrying quantum properties. Matan \textit{et al}. \cite{Tzur2025a} demonstrated that the quantum properties of BSV can be transferred to XUV through the absorption of one or two BSV photons, allowing the reconstruction of the quantum state and fluctuations of the emitted radiation at the single-atom level in such quantum two-color field. However, the macroscopic propagation of such quantum-laser-driven HHG in strongly driven dense media remains largely unexplored. In contrast to the well-understood propagation effects in coherently driven HHG, BSV fields exhibit large photon-number fluctuations that may significantly modify the coherence and quantum properties during propagation\cite{Javier2025Arxiv}. Moreover, macroscopic propagation in dense gases plays a critical role in the generation of intense XUV radiation. Understanding these effects is therefore essential for producing bright quantum XUV sources with attosecond pulse durations, and for advancing attosecond quantum spectroscopy\cite{Mor2026NP_app}.
	
	In this work, we present a combined experimental and theoretical study of macroscopic propagation effects in quantum high-order harmonic generation (QHHG), where a two-color quantum driving field—comprising a strong 800-nm coherent laser and a weak 1600-nm BSV field—interacts with krypton (Kr) atoms. Both the experimental measurements and comprehensive theoretical simulations of macroscopic QHHG demonstrate that the optimal gas pressure for the yield of odd (H$_{2N+1}$), even (H$_{2N}$) and satellites (H$_{(2N+1)\pm \frac{1}{2}}$) components significantly differ from each other. Phase-matching analysis based on the wave-mixing model reveals that the perturbation of 1600-nm BSV field on action phase of electron wave packet modifies the phase mismatch for each sub burst. It consequently results in the gas-pressure dependent harmonic yields and contrast between these components. The shot-to-shot harmonic spectral intensity and time domain second-order correlation function $g^2(0)$ extracted from numerical simulations confirm their quantum property of fluctuation. Our work demonstrates the generation and potential macroscopic manipulation of bright quantum XUV sources, which is important for quantum spectroscopy.
	
	\section*{Results}\label{sec2}
	\subsection*{Generation of bright quantum high-order harmonics}\label{sec2:1}
	
\begin{figure}[ht!]
		\begin{center}
			\includegraphics[width=0.95\linewidth]{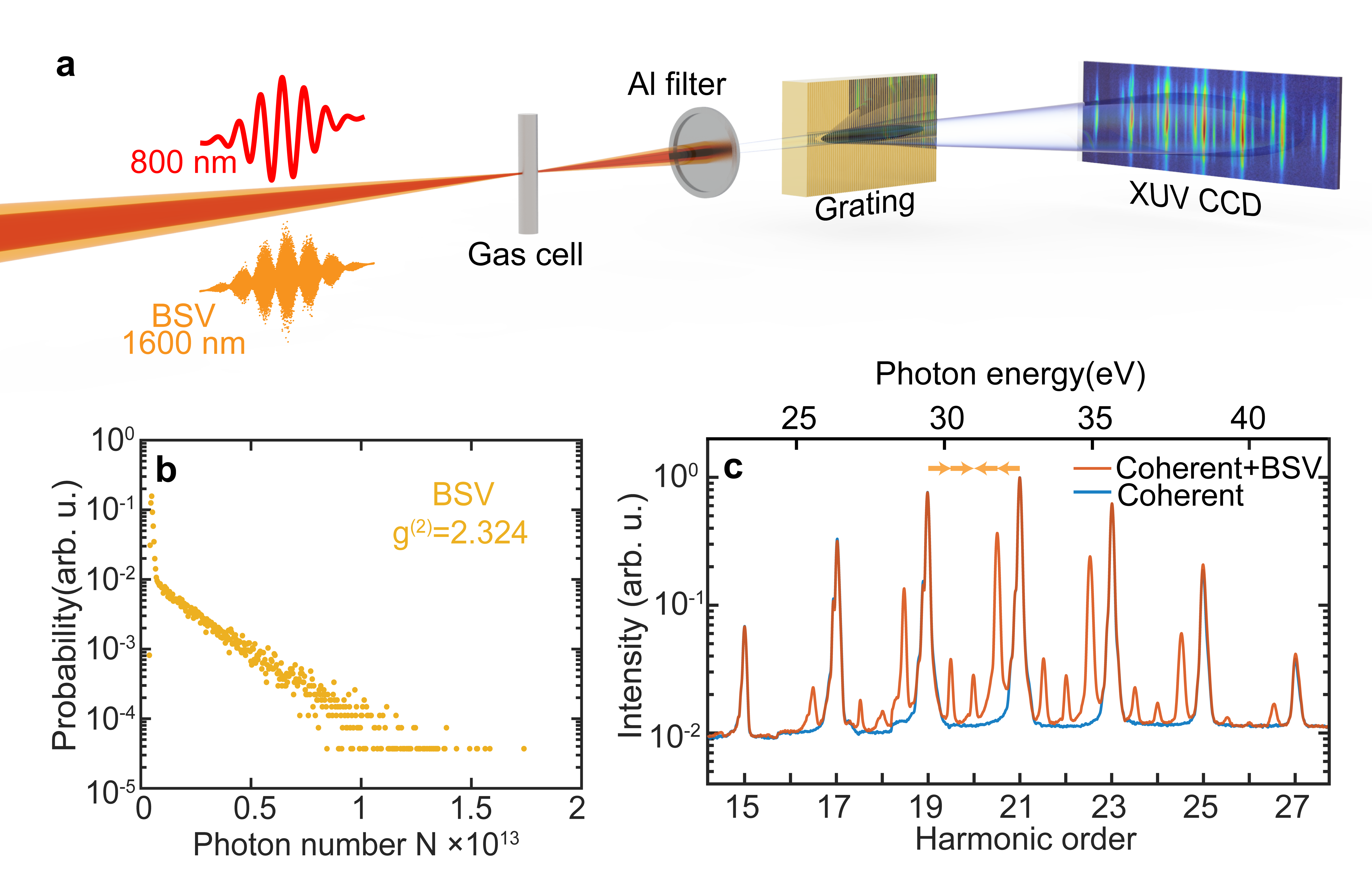}
		\end{center}
		\caption{\textbf{Schematic of quantum high-order harmonic generation with hybrid coherent and BSV light.}
			\textbf{a} Schematic of the HHG beamline driven by a hybrid field composed of a coherent laser (800 nm, red line) and single-spatial-mode BSV (1600 nm, orange dots).
			\textbf{b} Experimentally measured intensity distribution of the BSV field. The time domain second-order correlation $g^2(0)$ is 2.324. 
			\textbf{c} Integrated HHG intensity driven by a coherent laser only (blue line) and by a combined coherent+BSV field (red line) in Kr gas at the gas pressure of 42 Torr. The laser intensity is approximately $1.7 \times 10^{14}~\mathrm{W/cm^2}$ for the coherent field, and that of the BSV field reaches 2 $\times$ 10$^{11}$ W/cm$^2$.
		}
		\label{fig:schematic}
	\end{figure}

	In the experiment, as illustrated in Fig.~\ref{fig:schematic}a, a 35-fs, 800-nm laser pulse is split into two arms. Approximately $50\%$ of the pulse energy forms the coherent driving field, while the remaining energy is directed to the BSV generation arm. The BSV field is produced via spontaneous parametric down-conversion (SPDC) in a 3-mm type-I $\beta$-barium borate (BBO) crystal. Then the synthesized quantum two-color field interacts with Kr gas contained in a 3-mm gas cell. More experimental details are provided in the Methods section and Supplementary Information (SI). The quantum fluctuations and second-order correlation function $g^2(\tau = 0) = \langle N^2\rangle/\langle N \rangle^2$ are characterized and presented in Fig.~\ref{fig:schematic}b, yielding $g^2(0) = 2.324$ for the BSV field, whereas the coherent laser exhibits $g^2(0) = 1.003$ and its relative intensity distribution is presented in the SI. A motorized delay stage in the BSV arm controls the temporal overlap of the two fields, while their spatial overlap is monitored using cameras placed after recombination and near the focusing region. The gas pressure in the cell is precisely controlled and monitored (MKS Instruments, 640B) to study the phase match of the QHHG process. The emitted harmonics are recorded using a home-built XUV spectrometer equipped with an X-ray CCD camera (PIXIS-XO: 400B). Further details of the XUV spectrometer and gas cell can be found in our previous works\cite{Li2023,li2025Light}.
	
	The measured harmonic spectra, both with and without the BSV field, are presented in Fig.~\ref{fig:schematic}c, allowing a direct comparison of how the BSV field modifies high-harmonic generation. In the absence of the BSV field, only the dominant odd harmonics (H$_{2N+1}$) are observed, originating from the interaction of the coherent 800-nm field with Kr gas. When the BSV field is introduced, new spectral features appear: half-integer harmonics arise from single-BSV-photon processes, corresponding to the additional absorption or emission of one BSV photon from the main harmonics leading to (H$_{(2N+1)\pm\frac{1}{2}}$) components, while the even-order harmonics (H$_{2N}$) result from the absorption and emission of two BSV photons from the main harmonics, as illustrated by the orange arrows.
	
	\subsection*{Pressure-dependent yield of quantum high-order harmonics}\label{sec2:2}
	
\begin{figure*}[htbp]
		\begin{center}
			\includegraphics[width=1\linewidth]{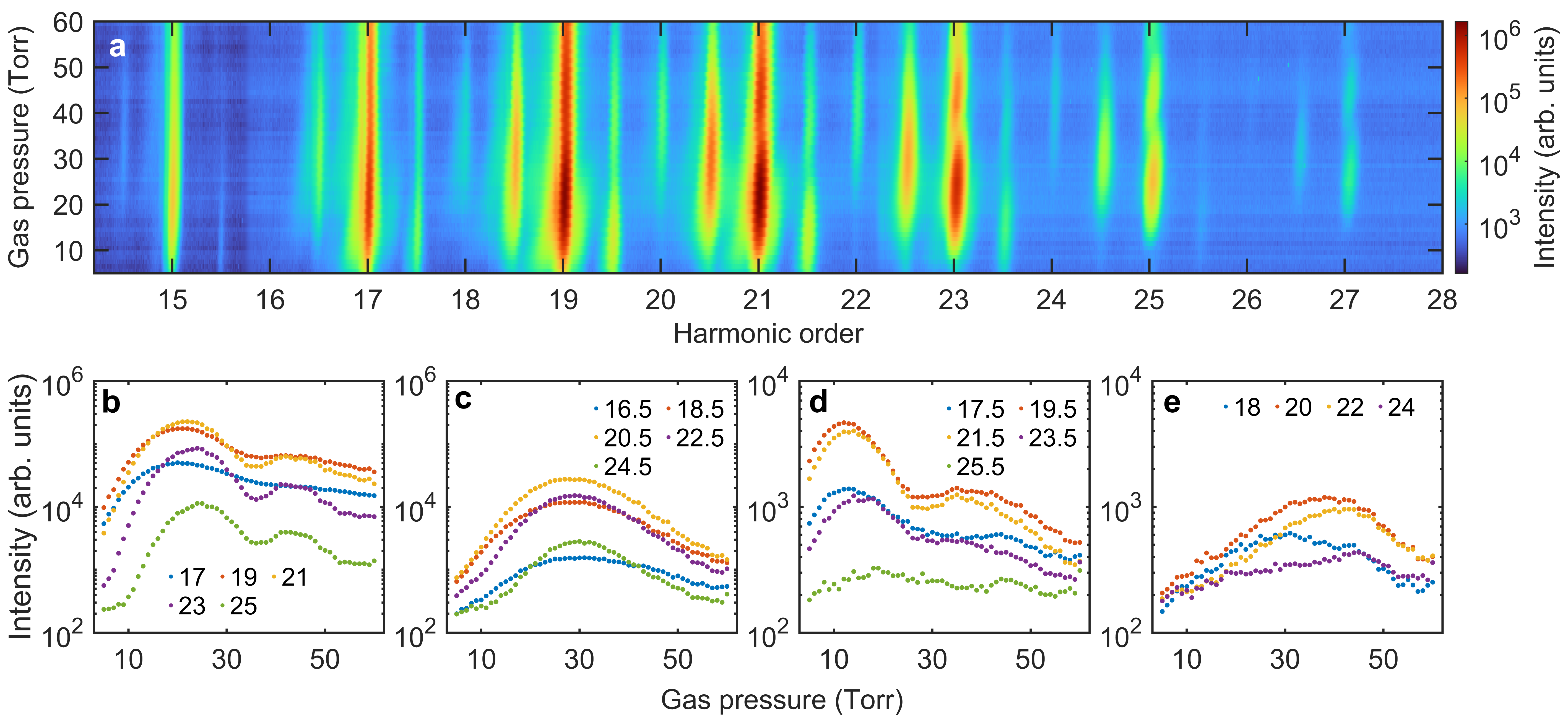}
		\end{center}
		\caption{\textbf{Experimentally measured high-order harmonics spectra.}
			\textbf{a} Experimentally measured HHG spectra under varying gas pressure from 5 to 60 Torr, with an 800-nm ($1.7\times10^{14}~\text{W}/\text{cm}^2$) and a 1600-nm BSV light ($2\times10^{11}~\text{W}/\text{cm}^2$). The integrated intensities of \textbf{b} main harmonics (H$_{2N+1}$), \textbf{c, d} satellite harmonics (H$_{(2N+1)\pm \frac{1}{2}}$) with single-BSV-photon emission and absorption, and \textbf{e} even harmonics (H$_{2N}$) respectively.
		}
		\label{fig:presssusre_scan}
	\end{figure*}

	Figure~\ref{fig:presssusre_scan} presents the pressure-dependent harmonic intensity generated in the quantum two-color field for harmonic orders spanning from H${14.5}$ to H${27}$, corresponding to the photon energies of approximately 22.5–42 eV. Notably, these components exhibit distinct gas-pressure dependence, indicating different phase-matching conditions. To highlight this behavior, we plot the pressure-dependent intensities of four components, H$_{2N+1}$, H$_{2N+\frac{1}{2}}$, H$_{2N+\frac{3}{2}}$, and H$_{2N}$, and show them in Fig.~\ref{fig:presssusre_scan}(b-e). 
	Within each component, all harmonic orders display similar trends, and their optimal gas pressures lie within a narrow range, showing weak dependence on the harmonic order. In contrast, significant differences are observed between the components. Specifically, the optimal gas pressure of the H$_{2N+1}$ harmonics is around 20 to 24 Torr, while those pressures for H$_{2N+\frac{1}{2}}$ and H$_{2N+\frac{3}{2}}$ are higher and lower, i.e., 27 Torr for H${20.5}$ and 13 Torr for H${21.5}$, respectively. And the phase-matching pressure for H$_{2N}$ lies around 30 to 44 Torr. A comparison between the pressure dependence of H$_{2N+1}$ in the quantum two-color field and that obtained using the 800-nm coherent field alone shows nearly identical behavior (see Fig. S2 in SI), confirming that the BSV field remains within the perturbative regime. However, the remarkable difference in the pressure dependence of the half-integer and even harmonics implies a novel phase-matching mechanism of QHHG. This abnormal pressure dependence of H$_{2N+\frac{1}{2}}$, H$_{2N+\frac{3}{2}}$ and H$_{2N}$ allows us to generate bright quantum light in the XUV regime and to control the relative intensity of half-integer and even harmonics involving absorbing and emitting BSV photons, with promising implications for future attosecond quantum spectroscopy and its applications\cite{Mor2026NP_app}. 
	
	\subsection*{Macroscopic propagation simulation of high-order harmonics under quantum fluctuation}\label{sec2:3}

\begin{figure}[ht!]
		\begin{center}
			\includegraphics[width=1\linewidth]{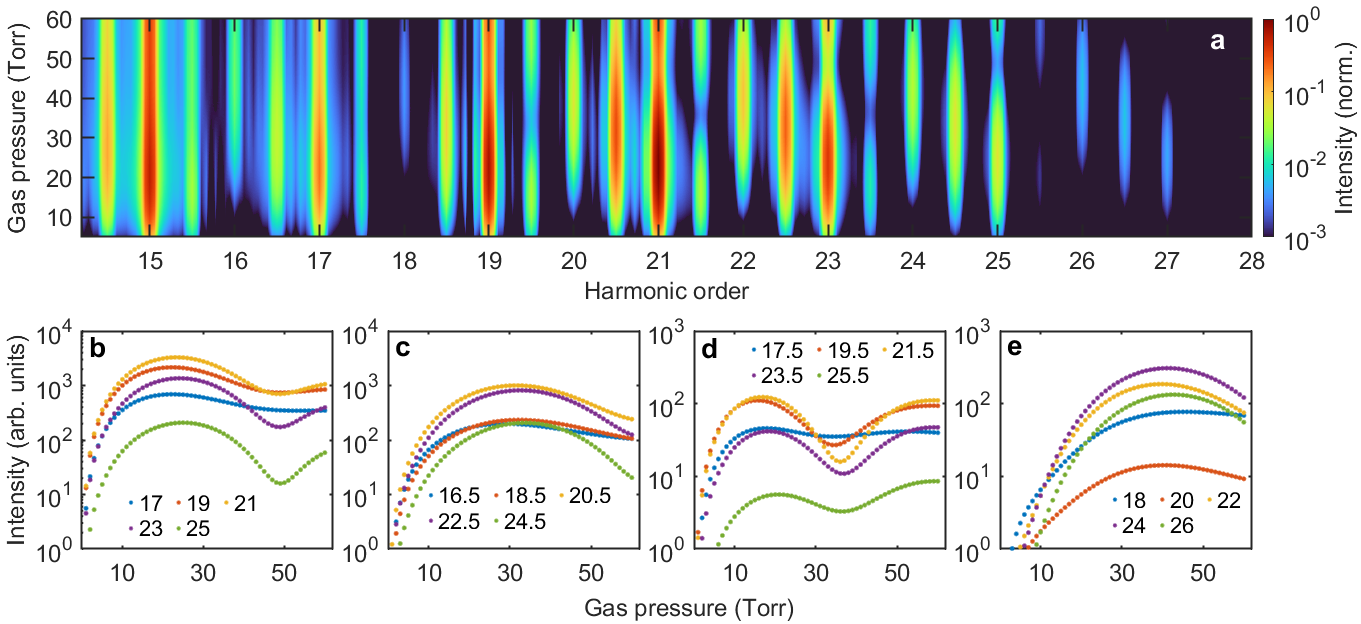}
		\end{center}
		\caption{\textbf{Simulated macroscopic high-order harmonic spectra.}
			\textbf{a} High-order harmonic spectra after macroscopic propagation under different gas pressures. The peak intensity of the coherent 800-nm field is $1.7 \times 10^{14}$ W/$\text{cm}^2$. And the 1600-nm BSV field is simulated by an ensemble of coherent fields containing 1000 samples described by a Husimi distribution. The harmonic intensities under varying gas pressure are shown for the \textbf{b} odd ($H_{2N+1}$), \textbf{c, d} satellites ($H_{2N \pm \frac{1}{2}}$) and \textbf{e} even ($H_{2N}$) harmonics.
		}
		\label{fig:sim_spec}
	\end{figure}

	The simulated HHG spectra under the experimental conditions are shown in Fig.~\ref{fig:sim_spec}. It nicely reproduces the experimental measurements, validating the theoretical propagation model. It is obvious to see that the best phase-matching pressure for H$_{2N+\frac{1}{2}}$ is around 30 Torr, which is higher than that of H$_{2N+\frac{3}{2}}$. Meanwhile, the H$_{2N}$ component requires a much higher pressure, around 40 Torr, to achieve the phase-matching condition. The behaviors of the harmonics along with varying gas pressures, as well as the optimal gas pressure with respect to each component, are well consistent with the experimental observations shown in Fig.~\ref{fig:presssusre_scan}. 

\begin{figure}[h!]
		\begin{center}
			\includegraphics[width=0.9\linewidth]{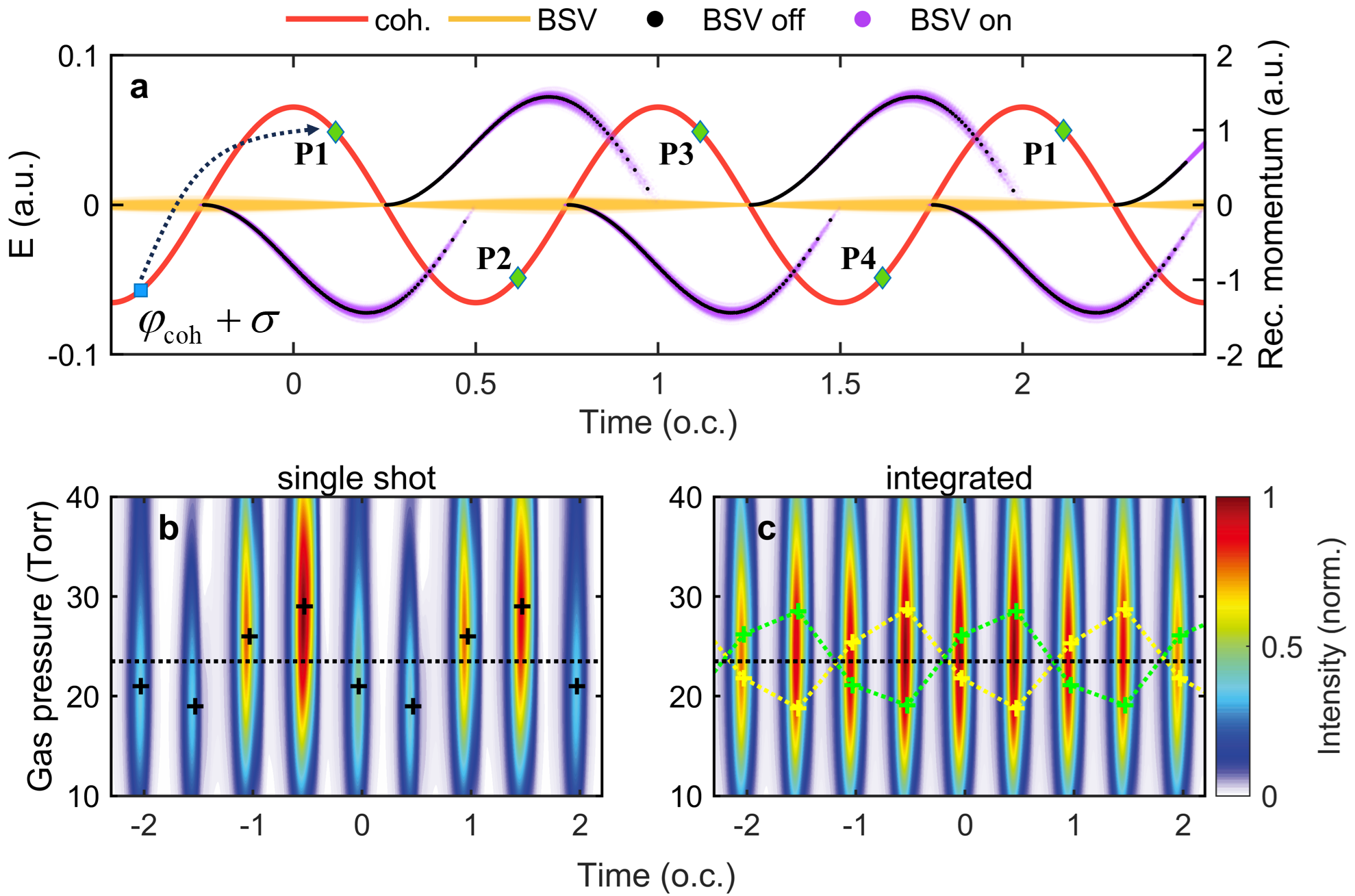}
		\end{center}
		\caption{\textbf{Phase-matching mechanism based on the wave-mixing model. }
			\textbf{a} Classical depiction of harmonic emissions. The weak BSV field introduces an extra phase \(\sigma\) to the action phase accumulated in 800-nm field only $\varphi_{\mathrm{coh}}$, and four individual emissions labeled by P1–P4 jointly contribute to the final HHG spectrum. \textbf{b} Simulated gas-pressure dependence of single-shot attosecond pulse trains from H16–H28 harmonics, with 1600-nm field amplitude of 0.0025 a.u. and phase of 0 rad. Black crosses mark the optimal gas pressure for each burst, and dotted line represents the optimal pressure driven by the 800-nm field only. \textbf{c} Same as \textbf{b} but for integrating all shots. Yellow and green crosses indicate the optimal gas pressure of two individual shots with identical amplitude but phases of 0 and $\pi$ rad, respectively.
			\label{fig:fwm_model}}
	\end{figure}

	Based on the consistency between the experimental results and the simulations employing the Husimi approximation, we establish a wave-mixing model to analyze the phase-matching mechanism. As illustrated in Fig.~\ref{fig:fwm_model}a, due to the broken temporal symmetry, four contributions with the same recombination kinetic energy (labeled by P1 - P4) are resolved in every two optical cycles of the coherent field. Solving the classical Newton equation for electron motion in the driving field reveals that the perturbation of the recombination time is marginal. In contrast, the action phase $\phi_{\mathrm{act}}$, defined as the temporal integral of the canonical Hamiltonian, differs substantially across these four contributions. By decomposing the action phase into the sum of that in the coherent field $\varphi_{\mathrm{coh}}$ and a perturbation term $\sigma$, we find that $\phi_{\mathrm{act}}$ is periodically modulated with the relative phase between the two fields and scales proportionally to the amplitude of the 1600-nm field\cite{Tzur2025a} (See Fig. S5 of SI).

	In the perspective of phase matching, there are four terms that contribute to the mismatch\cite{Weissenbilder2022}
	\begin{equation}
		\Delta  k = \Delta k_{\mathrm{geo}} + \Delta k_{\mathrm{dis}}+\Delta k_{\mathrm{ele}}+\Delta k_{\mathrm{dip}},
	\end{equation}
	where $\Delta k_{\mathrm{geo}}$ is resulted from the focal phase, $\Delta k_{\mathrm{dis}}$ and $\Delta k_{\mathrm{ele}}$ are caused by the change of refractive index due to neutral atoms and ionized electrons, which depends on the degree of ionization. $\Delta k_{\mathrm{dip}}=\nabla \phi_{\mathrm{act}}$ is related to the action phase. In our study, the BSV field is too weak to significantly alter the ionization degree, as evidenced by the pressure-dependent HHG measurements shown in Fig. S2 of SI, indicating that both $\Delta k_{\mathrm{dis}}$ and $\Delta k_{\mathrm{ele}}$ are expected to remain unchanged compared to the case of the coherent field alone. The same holds for $\Delta k_{\mathrm{geo}}$, except that the different focusing geometries of the two fields introduce an additional relative phase. Consequently, the difference in phase-matching conditions with and without the BSV field primarily arises from the action phase. 
	Using the relation $\phi_{\mathrm{act}}=\varphi_{\mathrm{coh}}+\sigma$, the phase mismatch $\Delta k$ can be further written as
	\begin{equation}\label{Eq:dk_purt}
		\Delta k = \Delta k_{\mathrm{coh}} + \nabla \sigma,
	\end{equation}
	where $\Delta k_{\mathrm{coh}} = \Delta k_{\mathrm{geo}}+\Delta k_{\mathrm{dis}}+\Delta k_{\mathrm{ele}}+\nabla\varphi_{\mathrm{coh}}$ represents the phase mismatch when the coherent field is present alone. When the perturbation field exists, an extra gas pressure $\Delta p$ is required to compensate this additional phase mismatch. Then the optimal gas pressure can be written as $p_{\mathrm{opt}} = p_{\mathrm{coh}}+\Delta p$, with the phase-matching gas pressure $p_{\mathrm{coh}}$ in the sole coherent field, and
	\begin{equation}
		\Delta p \cdot \frac{\omega_{\mathrm{h}}}{2\varepsilon_0ck_BT}\cdot\left[(1-\eta)(\alpha_1 - \alpha_q)-\frac{\eta e^2}{m_e \omega^2_{1}} \right] = -\nabla \sigma.
	\end{equation}
	Here, $\hbar\omega_{\mathrm{h}}$ is harmonic photon energy, $\varepsilon_0$ and $c$ are permittivity and light velocity in vacuum. $k_B$ and $T$ are Boltzmann constant and temperature. $\eta$ is the ionization degree of medium. $\alpha_1$ and $\alpha_q$ are polarizability of atom at fundamental and harmonic frequency respectively. $e$ and $m_e$ are charge and mass of electron, and $\omega_1$ is angular frequency of the coherent field. The two terms in square brackets represent the change of refractive index by neutral atoms and plasma, respectively. Thus, the optimal gas pressure of each contribution P1 - P4 are distinguished by their perturbation phases $\sigma$. As shown in Fig.~\ref{fig:fwm_model}b, the optimal gas pressure for achieving the maximum burst intensity of each contribution is offset from the phase matching gas pressure (indicated by black dotted line) when using only coherent field. The offsets for P1 and P3 (P2 and P4) are opposite in sign, following the relation $\sigma_3 = -\sigma_1, \sigma_4 = -\sigma_2$ \cite{Tzur2025a}. These offsets also depend on the amplitude of the 1600-nm field (See Fig. S5 of SI), qualitatively consistent with Eq.~(\ref{Eq:dk_purt}). In Fig.~\ref{fig:fwm_model}c, integration of all samples, i.e., the expectation of pulse intensity in the BSV field, exhibits a pressure range for optimal QHHG yield similar to that of the coherent laser (See Fig. S4 of SI). This visual similarity is directly related to the feature of 0 - $\pi$ phase ambiguity characteristic of BSV fields, as illustrated by the colored lines in Fig.~\ref{fig:fwm_model}c.
	
	\subsection*{Pressure-dependent fluctuation of harmonics}\label{sec2:4}

\begin{figure}[ht!]
		\begin{center}
			\includegraphics[width=1\linewidth]{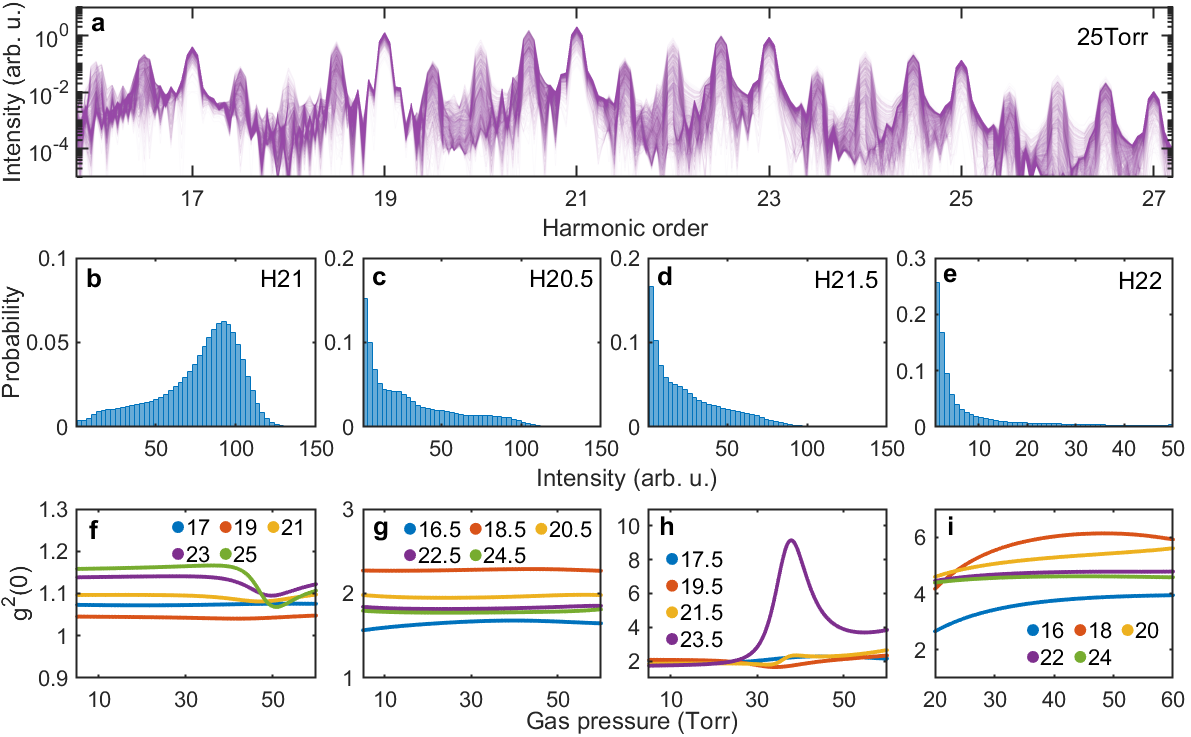}
		\end{center}
		\caption{\textbf{Simulated macroscopic high-order harmonic spectra and photon statistics driven by hybrid coherent and BSV light.}
			\textbf{a} Shot-to-shot high-order harmonic spectra after macroscopic propagation at 25 Torr. 
			Photon statistics of harmonics at 25 Torr for \textbf{b} H21, \textbf{c} H20.5, \textbf{d} H21.5 and \textbf{e} H22.
			The variation of time domain second-order correlation $g^2(0)$ along with gas pressure corresponds to \textbf{f} odd ($H_{2N+1}$), \textbf{g, h} satellites ($H_{2N \pm \frac{1}{2}}$) and \textbf{i} even ($H_{2N}$) harmonics.
			\label{fig:spec_shot_fluc} 
		}
	\end{figure}

    Based on the wave-mixing model presented above, the final harmonic spectra arise from the coherent superposition of four distinct emission pathways. A crucial point is that, since the sub-cycle phase-matching conditions depend on the perturbative quantum field, the phase matching can fluctuate from shot to shot even under identical macroscopic conditions, due to the varying amplitude of the BSV field. Consequently, it is not obvious that the quantum fluctuations of the harmonics will remain identical to those of the single-atom response. Figure~\ref{fig:spec_shot_fluc}a displays the simulated shot-to-shot harmonic spectra following macroscopic propagation at 25 Torr. At this pressure, the H$_{2N+\frac{1}{2}}$ harmonics are better phase-matched than the H$_{2N+\frac{3}{2}}$ ones, and the intensity of even-order harmonics are also pronounced. The photon statistics of the four harmonic components are presented in Fig.~\ref{fig:spec_shot_fluc}(b–e). The H21 harmonic exhibits a Gaussian-like distribution, closely resembling that driven by a coherent field. In contrast, H20.5, H21.5, and H22 show probability distributions that decay slowly with increasing photon number, revealing bunching and super-bunching characteristics—a trend reminiscent of the BSV field.
    
    Figure~\ref{fig:spec_shot_fluc}(f–i) shows the pressure-dependent second-order correlation, $g^2(0) = \langle I^2 \rangle/\langle I \rangle^2$, for the four harmonic components. A clear pressure-dependent variation of $g^2(0)$ is observed for all four components, and within each component, different harmonics exhibit distinct intensity fluctuations. Notably, when a harmonic is not well phase-matched, $g^2(0)$, can change significantly. For instance, as the gas pressure increases from 40 Torr to 50 Torr—a region where the harmonics are phase-mismatched, the $g^2(0)$ drops rapidly for H21, H23, and H25 starting at 40 Torr, then recovers after 50 Torr. A similar effect occurs near 30–40 Torr for H19.5, H21.5, and H23.5, which also fall in a phase-mismatching regime. Moreover, for H$_{2N}$ harmonics, the pressure dependence of $g^2(0)$ is more pronounced than for the other components. In the phase-mismatching case, the shot-to-shot intensity fluctuations are influenced not only by the driving laser fluctuations but also—more significantly—by macroscopic propagation effects in the dense medium. Although $g^2(0)$ is sensitive to gas pressure, its values remain approximately 1.1 for H$_{2N+1}$, around 2 for H$_{2N \pm \frac{1}{2}}$, and exceed 4 for H$_{2N}$ near the phase-matched pressure. This behavior is consistent with reported measurements \cite{Tzur2025a}, confirming that the quantum fluctuations of the harmonics are effectively transferred from the driving field and validating the bright QHHG scheme.

	\section*{Discussion}\label{sec3}
	In conclusion, we have experimentally and theoretically investigated the macroscopic propagation effect of quantum high-order harmonics using the combination of a strong coherent field and a weak BSV field. The harmonic components of H$_{2N+1}$, H$_{2N+\frac{1}{2}}$, H$_{2N+\frac{3}{2}}$, and H$_{2N}$ exhibit distinctly different dependencies on gas pressure in terms of their yield. Numerical calculations show that the BSV field perturbs the action phase of the subcycle emission bursts, leading to different phase-mismatch behaviors and thereby modifying the macroscopic QHHG spectral structure as the gas pressure varies. This effect corresponds to a modulation of harmonic intensity fluctuations, photon bunching, as well as the relative intensity between these components. Meanwhile, our findings demonstrate the significant potential for spectral, temporal, and quantum properties control of bright attosecond quantum light sources by tuning macroscopic phase-matching conditions. This capability opens new directions for attosecond quantum spectroscopy and its application to probing quantum electron dynamics in matter.
	
	\section*{Methods}\label{sec4}
    \subsection*{Preparation of quantum two-color fields}
    The Ti: sapphire laser (800 nm, 35 fs, 1 kHz) is split into two paths (BS2, 50:50) for preparing the coherent and BSV light. In the BSV path, the reflected coherent beam passes through a telescope (4:1) to reduce the beam size to $\sim$ 2~mm and enhance the BSV generation efficiency. The BSV field is generated in a reflection-based geometry through a double pass in a 3-mm-thick type-I $\beta$-BaB$_2$O$_4$ (BBO) crystal (cut-angle, 19.8$\degree$). A 0$\degree$ reflection silver mirror (35~cm behind the crystal) is mounted on the second delay line to optimize the spatial mode toward a single-mode profile. The generated BSV is then decoupled from the input coherent laser by the dichroic mirror (DM), and then directed through a second telescope (1:7) to expand its beam size to $\sim$ 14~mm. After passing a 1600~nm bandpass filter, the BSV has $g^{(2)} = 2.324$ with the power of 0.18~$\mu$J ($\sim 1.5 \times 10^{12}$ photons). The coherent and BSV beams are then recombined via an off-axis holey parabolic mirror and sent through krypton in a 3~mm gas cell for high-harmonic generation. 

    \subsection*{Theoretical model}
	Although the BSV field has a distinct quantum fluctuation against the coherent field, any single realization of the pulse has a definite waveform\cite{Kern2026}. This allows us to examine the single-shot propagation process based on the well-established theory of HHG by a coherent field\cite{Jin2018,Tang2024}. To elucidate this intriguing behavior, we perform theoretical simulations on macroscopic propagation of QHHG by solving the one-dimensional time-dependent Schrödinger equation (1D-TDSE) coupled with Maxwell equations involving macroscopic propagation effects. The driving field was modeled as a superposition of a noiseless fundamental component at $\omega$ (800 nm), and a BSV field at $0.5~\omega$ (1600 nm) characterized by Husimi distribution $Q(\alpha)=\pi^{-1}\bra{\alpha}\hat{\rho}\ket{\alpha}$ with density matrix element operator $\hat{\rho}$ and coherent state $\ket{\alpha}$ \cite{Tzur2025a}.
	According to the experimental condition, the 800-nm spatial field is analytically expressed as the fundamental Gaussian mode, with peak intensity of $1.7\times10^{14}~\text{W}/\text{cm}^2$ and beam waist of $100~\mu\mathrm{m}$. And the spatial mode of 1600-nm field is also the fundamental Gaussian mode with beam waist of $25~\mu\mathrm{m}$. Then, 1000 shots satisfying the Husimi distribution are random sampled for BSV statistical simulation (See Fig. S3 of SI). To make this computationally tractable, the 1D-TDSE calculations were significantly accelerated using a machine-learning-based approach. 
	The macroscopic propagation of harmonics was performed by inserting the polarization obtained from solving 1D-TDSE into the propagation equation that incorporates the dispersion and absorption of gas medium. In simulations, the entrance plane of the 3-mm gas cell of Kr atoms is 3.5 mm before the focus, and phase modulation of driving field by neutral atoms and plasma is considered. More details of the numerical implementation can be found in the SI.
	

	\backmatter

	\bmhead{Acknowledgments}
	The authors thank Prof. Anne L'Huillier, Prof. Cheng Jin, Prof. Xu Wang, and Dr. Haoyu Liu for helpful discussions. This work was supported by National Natural Science Foundation of China (grants no. 12450402, 12450404, and 12134005).
	
	\bmhead{Author contributions}
	W.W., L.W., L.H., M.L. and S.L. conducted the experiments; Y.S., X.T. and J.Y. conducted the simulations; J.Y., and S.L. supervised the project; W.W., Y.S., X.T., M.L., D.D., J.Y. and S.L interpreted the data and contributed to the preparation of the manuscript. All authors discussed and approved the results and the manuscript.

    \bmhead{Data availability}
    The source data that support the findings of this study are available in \textit{figshare} platform (link: \href{https://figshare.com/s/9c811ab5777abbbbdfb5}{https://figshare.com/s/9c811ab5777abbbbdfb5}).
    
	\bmhead{Conflict of interests}
	The authors declare no competing interests.
	
	\bmhead{Supplementary information}

	The Supplementary Information text file contains details about experimental setup and method, HHG from the coherent field and quantum two-color field, and theoretical methods.

	
\end{document}